\documentstyle[12pt,aasms4]{article}

\newcommand{\gsim}{${\raisebox{-.9ex}{$\stackrel{\textstyle>}{\sim}$}}$ }
\newcommand{\lsim}{${\raisebox{-.9ex}{$\stackrel{\textstyle<}{\sim}$}}$ }
\newcommand{\avg}[1]{\langle #1 \rangle}

\begin{document}

\title{Lifetime of Surface Features and Stellar Rotation:\\
    A Wavelet Time-Frequency Approach}

\author{Willie Soon\altaffilmark{1}, Peter Frick\altaffilmark{2}, Sallie Baliunas\altaffilmark{1}}

\altaffiltext{1}{Harvard-Smithsonian Center for Astrophysics, 
60 Garden Street, Cambridge, MA 02138} 
\altaffiltext{2}{Institute of Continuous Media Mechanics,
Korolyov str.1, 614061 Perm, Russia}

\begin{abstract}
 We explore subtle variations in disk-integrated measurements spanning
 $\lsim$ 18 years of stellar
 surface magnetism by using  a newly developed time-frequency gapped 
 wavelet algorithm.
 We present results based on analysis of the Mount Wilson  
 Ca II H and K emission fluxes in four,
 magnetically-active
 stars (HD 1835 [G2V], 82885 [G8IV-V], 149661 [K0V] and 190007 [K4V])
 and sensitivity
 tests using artificial data.

 When the wavelet basis is appropriately modified  (i.e.,
 when the time-frequency resolution is optimized),
 the results are consistent with the
 existence of spatially localized and long-lived 
  Ca II features (assumed here as activity regions that tend to
 recur in narrowly-confined latitude bands), 
 especially in HD 1835 and 82885.
 This interpretation is based on the observed
 persistence  of relatively localized Ca II wavelet power at a 
 narrow range of rotational time scales,
 enduring as long as $\gsim$  10 years.

\end{abstract}


\section{Introduction}

 At Mount Wilson Observatory (MWO) the Ca II H (396.8 nm)
 and K (393.4 nm) emission fluxes 
 have been observed as a proxy for surface magnetism
 since 1966 in many lower main sequence stars.
 Since 1980, observations have been
 obtained several times per week over periods as long
 as several months.
 These densely-sampled records are suited 
 for studying changes in the pattern
 of the Ca II chromospheric radiative losses on timescales of days to months
  (Baliunas et al. 1995).

 We present results from a new wavelet-type  analysis of the 
 Ca II records for  four 
 stars with higher mean Ca II emission fluxes
 and faster or comparable mean rotation than the Sun:
 HD 1835, 82885, 149661 and 190007 (Table 1).
 The records of those four stars
  were chosen as a starting point for complementary,
 upcoming studies, with E. Smith, A. Vaughan and colleagues
 at the Jet Propulsion Laboratory, that seek to understand 
 phenomena linked to persistent magnetic structures on the Sun.
  Theoretical results
 (Sch\"ussler et al. 1996; DeLuca, Fan \& 
Saar 1997) suggest that such active,  rapidly rotating stars
 should show magnetic flux tubes in limited latitude zones
  which may 
  facilitate the persistent clustering of surface features
 (see e.g., Seiden \& Wentzel 1996; Ruzmaikin 1998 for
  discussion on mechanisms associated with 
  the clustering of emerging magnetic fluxes on the Sun).

 These few stars  were 
 noted to as showing long-lived Ca II emitting regions in
 an earlier analysis of the then shorter 
 MWO records (Vaughan \& Baliunas 1992; see also Noyes 1982). 
 But the present wavelet analysis 
 constitutes two pragmatical improvements over the previous analysis.
 First, the length of the records has nearly doubled.
 Second, in contrast to the
 previous Fourier transform technique, 
  no apriori assumption needs to be made 
 concerning the Ca II rotational modulation with 
 time.
 Since this is an exploratory effort meant to examine
 the suitability of the technique
 we derive no information on how common or uncommon
 the results on long-lived surface Ca II features
 are among the stars in the entire MWO database.
 That task will be taken up in the future.

\section{Method of Analysis: The Gapped Wavelet Algorithm}

  Because of  mathematical refinements of the wavelet transform
 (i.e., self-similarity of the wavelet basis function,
  time-frequency localization),
  it  has become increasingly popular 
 as a tool for extracting local-frequency information
  (e.g., Farge 1992; Kumar \& 
Foufoula-Georgiou 1997).
 The wavelet transform differs from traditional Fourier
 analysis because of its ability to
  efficiently detect multi-scale,   non-stationary processes.

 We applied a newly introduced {\it gapped} wavelet algorithm\footnote{The algorithm was first introduced under the name {\it adaptive} wavelet
 but because this term has been widely used for different wavelet
 techniques, it has been renamed
  {\it gapped } wavelet (see Frick, Grossmann \& Tchamitchian 
 1998 for 
 mathematical motivations and proofs).}
 (Frick et al. 1997) 
 to the MWO records of four stars to study time variations 
 of rotational modulation of the Ca II fluxes.
 The algorithm alleviates two constraints 
 in stellar activity records that complicate
 traditional methods of period analysis:
 limited duration and sampling gaps.
 This goal is achieved by fine correction (adaptation) of the wavelet
 for a given time and frequency while still satisfying the admissibility
 condition (for which 
 the mean value of the wavelet must be zero: $\avg{\psi} = 0$).
 The admissibility condition can be broken 
 when the wavelet overlaps  data  gaps in 
 or the edges of data series.

  In this analysis we use
  the Morlet wavelet with an adjustable parameter, $\kappa$,
  which can be fine-tuned  to yield  optimal  resolutions
 of time and frequency:
 \begin{equation}
 \psi(t) = e^{-t{^2}/2\kappa^2}e^{i2\pi t} \hspace*{2ex}.
 \end{equation}
 In that case, the  resolutions of the wavelet 
 for a given characteristic time-scale, $T$, are: 
\begin{equation}
\delta t = c \kappa T,   \hspace{6ex} \delta \omega = {c \over  \kappa T },
\end{equation}
  where $c$ is a constant of order unity.
  Small values of $\kappa$ give better time resolution 
 while large values of $\kappa$ improve frequency resolution.
  The key step in any application of the wavelet transform is
  to deduce the optimum trade-off between
 frequency and time resolution and the art is
 to choose $\kappa$ carefully so that it fits the  
 physical phenomena of interest.

 The commonly adopted value of $\kappa$ is 1; the 
 limit $\kappa \rightarrow \infty$ corresponds to the Fourier transform.
  But the choice of $\kappa$ is highly 
 restricted by the admissibility condition,  $\avg{\psi}=0$,
  so that  only a finite range and discrete values of $\kappa$ are allowed.
  Under the application of the standard Morlet wavelet algorithm,
  the admissibility condition breaks down
  when $\kappa$ is below unity because
  the real parts of the Morlet wavelet do
  not vanish.
 In contrast, because of the renormalization performed in
 the gapped wavelet algorithm,
 that problem is avoided automatically 
 and a wide range of  $\kappa$  (provided $\kappa$ is
 not too small) can be used.

\section{Results}

 We examine two aspects of stellar variability in the records:
 (I) the presence of  any dominant surface  magnetic 
     features and signatures of rotating features;
 (II) the nature of time variations 
      of any dominant rotational signals. 
 Results are presented for a wide range of
 $\kappa$ and illustrate that aspect (I) is best studied
 with lower values of $\kappa$ 
 while   (II) is more optimally addressed
 with larger values.

\subsection{Optimum time-frequency resolutions and beat phenomena: Tests from simple oscillator models}

 The MWO time series of Ca II emission fluxes  
 are assumed to be a 
 superposition of short-lived oscillations with random, 
 initial phases
 and individual frequencies,
 which have a mean value, $\Omega$ (or period, assumed to be rotation, $T=2\pi/\Omega$) 
 and characteristic width, $\Delta \Omega$.
 The choice for time-frequency resolution is set 
 by two factors: the typical time of an individual oscillation and 
 the band width, $\Delta \Omega$.
 If one is interested in studying variations
 in frequencies caused by the emergence of fluxes
 drifting in latitude,
 a wavelet with $\delta t$ on the order of typical lifetime
 of individual features should be used. 

  From a detailed analysis of
 the magnetic flux budget on the Sun,
 Schrijver \& Harvey (1994) deduced 
 that the surface-averaged lifetime of photospheric magnetic flux
 (which included contributions from magnetic bipolar and plage regions
 and other smaller-scale features)
 varies from $\sim$ 130 days at cycle maximum to
 340 days  at cycle minimum. 
 These timescales yield estimates of $\kappa$ from
 about 4 to 10. 
 For the Ca II records of young and active stars,
 the values for $\kappa$ may need to be smaller. 
 But small values of $\kappa$ produce frequency beating 
 because of reduced wavelet spectral/frequency resolution
 (or increased time resolution).
 We illustrate this phenomenon with two  examples. 
 First, we consider the simple superposition of two harmonic oscillators
 with nearby frequencies $\Omega$ and $\Omega + \Delta\Omega$. 
 Beating arises if the wavelet frequency resolution, 
 $\delta \omega$  $\gsim$ $\Delta\Omega$.
 In Figure 1a, the  wavelet power 
 with $\kappa=1$ is shown
 for the case of two superposed sinusoids with periods 9 and 11 days.
 The maximum wavelet power for time $t$ 
 displays a beating with
 characteristic frequency,  $\Omega_b = \Delta\Omega$.

 Figure 1b shows the modulus of wavelet coefficients for
 another artificial signal, one 
 which is produced by the superposition of
 short-lived variations with
 random phases and frequencies (9 $\lsim$ $T$ $\lsim$ 11 days).
 Again,   the local maxima in wavelet power for each time
 oscillates around the mean value
 of the dominant periods.
  The period of the
  beating yields an estimate of the average width of a band of frequencies.
 The beat frequency, $\Omega_b$, suggests
  $\Delta T = \Omega_b T^2/2 \pi  \approx 1 $ day,
  which represents the mean difference
 between the periods of any two
 individual oscillations.

 Figure 1c shows the results
 from the artificial series of Figure 1b 
 analyzed with a larger value of $\kappa=10.0$.
 The beat phenomenon disappears, and only 
  discrete periods are resolved at
 a given time interval.
  Thus, appropriate choices of $\kappa$
 allow different kinds of information to be extracted. 

\subsection{Ca II H and K Wavelet Spectra: HD 1835, 82885, 149661, 190007} 

 Figure 2 shows the time-averaged wavelet spectra for the four active stars.
 Table 1 summarizes the values of rotation periods determined from
 the gapped wavelet technique.
 The dominant timescales found here 
 correspond well with results from
 periodogram analysis 
(e.g., Donahue, Saar \& Baliunas 1996).

 To study the time variations of the rotation periods, 
 we started the analysis  with $\kappa =1$.
  We found significant 
  oscillations of the primary period
  in the time-frequency plane of the wavelet modulus for each of the records
 of the four stars.
 An example of these oscillations  is shown in Figure 3a for 
 HD 1835. 
 
 The maxima of wavelet power (traced by the filled circles) are themselves
 a time series, $T_m(t)$, from which a
 wavelet spectrum is also calculated.
 Those spectra for all four stars are shown in 
 Figure 3b.
 Each spectrum displays a pronounced
 peak corresponding to the main beat frequency, $\Omega_b$ 
  (Table 1).
 The test examples suggest that the beat
 frequency contains
 information on the range of unresolved  frequencies in each record.
 The dominant but unresolved periods range from one to a few days,
 and are also listed
 in Table 1. 
  The  beat frequency yields an
  estimate of the characteristic $\kappa^*$ (Table 1),
  and suggests  the choice of  a larger $\kappa$ that would be
  useful for studying, e.g.,
  the change in dominant periods over  extended times.  

 Figure 4 shows the time variations
 of the dominant timescales 
 for each star
 calculated with optimal wavelet time-frequency resolution.
  The values of the
  wavelet resolution parameter, $\kappa$, are 2.5, 2.0, 4.0, 
 and 4.0, respectively.
 These values were  chosen to be about
  2.5 times larger than
 $\kappa^{*}$ in order to:
 (i) avoid the beat phenomena shown in Figure 3
 and 
 (ii) investigate the relative long-term
     stability of the dominant peaks in the spectra.
 Figure 4 shows that
 the dominant rotational periods
 for  HD 1835, 82885 and 149661
 are relatively narrow in range 
 and remarkably stable throughout the full interval of observations.
 This pattern of time-frequency localization 
 may be interpreted as  
 recurring Ca II emissions from localized surface features
 that continued for $\gsim$ 10 years.
 The  dominant rotational peaks for HD 190007 seem less localized
  and it may or may not be dominated by narrowly distributed, 
  persistent  or recurring surface features.
  The range of each dominant scale is 
 assumed to be the wavelet frequency
  resolution, $\delta \omega$; the values for the records of the four stars
 are 0.5, 1.5, 0.5, 0.5 days, respectively.

\section{Discussion}

  Previous analyses of the MWO Ca II records used  Fourier methods
  for  studying surface rotation and
 pattern of surface differential rotation (e.g., Baliunas et al. 1985;
 Donahue et al. 1996).
 Two interconnected assumptions were made in those studies: 
 First, that by binning the data series into individual observing
  seasons ($\approx$ several months), time resolution for studying the
 underlying rotational modulation is automatically optimum.
 Second, once any dominant rotational signal is detected,
 systematic variations with time of the rotation period 
 can be studied by simply considering the change
 in rotational signal from one seasonal bin to another.
   Because the wavelet transform adopts 
 localized basis function,
 information on the time dependence of spectral properties is
   preserved, so it avoids the assumptions used in the Fourier studies. 
 However, although the wavelet method provides information on 
 spectral changes with time,
 there is a practical limit to the physical interpretation
 of the results because the spatial location of the stellar features is unknown.
 That means that proving that rotational time-scales associated
 with specific surface features are changing with time is difficult.

  On the other hand, one can invert the question and
  ask if a stationary, 
 rotational signal exists. This is an easier question to answer
  because the wavelet method can establish clearly the presence of
 a persistent signal. 
 The results in figure 4 suggest that the Ca II emitting regions marking
 the rotation periods last $\gsim$ 10 years, 
 for at least two of the four
 active stars analyzed.
 Thus, the idea of using the wavelet 
 time-frequency  approach to
 establish stationarity of the rotational signal  is 
 a useful strategy in gaining understanding of
 the surface Ca II activity.

  Because the Ca II features 
  in HD 1835 and 82885 seem to persist so much longer than
  lifetime of individual magnetic features (e.g., active regions) on the Sun,
  our results might caution against indiscriminate application of the
 stellar knowledge to  the Sun. 
 This would seem to be a conclusion opposing any hope  of 
  applying the solar-stellar connection in a two-way, quantitative manner.
 On the other hand, an analysis of sunspot groups from 
 1940 to 1956, suggests  that there are  
 recurrent clusters of sunspot nests  
 which can persist and keep 
 their rotation rate for up to several years,
 while at the same time showing systematic
 meridional drift towards equator with velocity of about 1 m/s
 (e.g., van Driel-Gesztelyi, van der Zalm \& Zwaan 1992).
 Those authors also noted that components in large active nests
 tend to overlap in time with their mean latitudes differing 
 by less than 2.5$^{^{\circ}}$ while 
 the difference in longitude may extend up to 55$^{^{\circ}}$.
 Thus, the observed persistence of Ca II features
 in our active star sample may not be entirely
 incompatible with characteristics observed on the Sun.

\acknowledgments

  We thank Cristina Christian and Robert Donahue for their
  valuable contributions  to the MWO HK project.
  We are also grateful for the dedicated efforts
  of our colleagues, especially Mike Bradford, Laura Woodard-Eklund,
  Jim Frazer  and  Kirk Palmer at Mount Wilson Observatory. 
  The comments by the referee led to improvement of
  this manuscript. 
  
  This work was supported by the  Electric Power Research Institute,
  Richard C. Lounsbery Foundation and MIT Space grant \# 5700000633. Additional support from the
 U. S. Civilian Research and Development Foundation Grant \# 171600
 is also gratefully acknowledged.
  This research was made possible by a collaborative agreement between the 
Carnegie  Institution of Washington and the Mount Wilson Institute. 

\clearpage

\begin{table*}

\tablenum{1}
\caption{Summary of characteristic time-scales}

\begin{center}

\vspace*{4ex}
\begin{tabular}{|c|c|c|c|c|c|}\tableline
 & $T_r$ & $T_b$ & $\Delta T_r$ & $\Delta T_r / {T_r} $ & $\kappa^{*}$ \\
\tableline
HD 1835 (G2V)  & 7.7 & 47  & 1.3 & 0.15 & 1 \\
HD 82885 (G8IV-V)  & 18  & 96  & 3.4 & 0.19 & 0.9 \\
HD 149661 (K0V) & 21  & 210 & 2.1 & 0.10 & 1.6 \\
HD 190007 (K4V) & 28  & 120-230 & 7-3 & 0.25-0.1  & 0.7-1.5 \\
\tableline
\end{tabular}
\end{center}

\tablecomments{$T_r$ is the rotation period deduced from time-averaged wavelet spectra (see Figure 2);
 $T_b$ is the beat period found from the time variations of the
        rotational time scales (see Figure 3);
 $\Delta T_r$ is the characteristic range of rotation periods
 deduced from values of the beat period interpreted as
 a simple model of oscillators with a range of $T_r$;
 $\kappa^{*}$ is the characteristic $\kappa$
 that distinguishes lower values for which the beat phenomenon
 is observed  from higher  values 
 for which time resolution becomes
  too poor to capture the beat among dominant periods.}

\end{table*}

\clearpage

\begin{center}
{\bf Figure Captions}
\end{center}

\figcaption{ Time-frequency plots of the
 modulus of the wavelet coefficients
 for artificial time series with:  (a) superposed oscillations with
       two ($T_1=9$ and $T_2=11$ days) distinct periods  ($\kappa =1$);
 (b) superposed, randomly-phased
  oscillations with multiple periods ranging from 9 to 11 days  ($\kappa =1$);
 (c) same as in (b) but calculated for $\kappa =10$.
 Lighter shading corresponds to increasing values of the
 wavelet modulus except that filled circles trace
  the local wavelet maxima.
 Figures (1b) and (1c) contrast
 the disappearance of the beat
 phenomenon when  $\kappa$ is
  increased. 
\label{fig1}}

\figcaption{ Time-averaged Ca II wavelet spectra, $W(T)$, for
 four stars: HD 1835, 82885, 149661 and 190007.
 Peaks in the spectra corresponding to rotation and 
  are located around 7.7, 18.0, 21.0 and
 28 days, respectively (Table 1). \label{fig2}}

\figcaption{(a) The modulus of Ca II wavelet coefficients for HD 1835
   calculated with $\kappa=1.0$  illustrates  
the fine-structure in the time variation
 of the primary rotation signal.
 Lighter shading indicates increasing values of the
 wavelet modulus
 (except for the  white regions of data gaps and the filled circles that trace
 the wavelet maxima).
  (b) The wavelet spectra of the time variations of the
        rotational time scales.
  The beat periods for the rotational scales 
  are  47, 96, 210 and 120-230 days for the four stars (see Table 1).
 \label{fig3}}

\figcaption{ Distribution of the modulus of Ca II wavelet coefficients,
  $W(T,t)$, for the four stars.
 Lighter shading corresponds 
 to increasing values of the
 wavelet modulus
 (except for the white regions of data gaps).
 The dominant scales of  Ca II wavelet power are
 traced by filled circles.
 \label{fig4}}

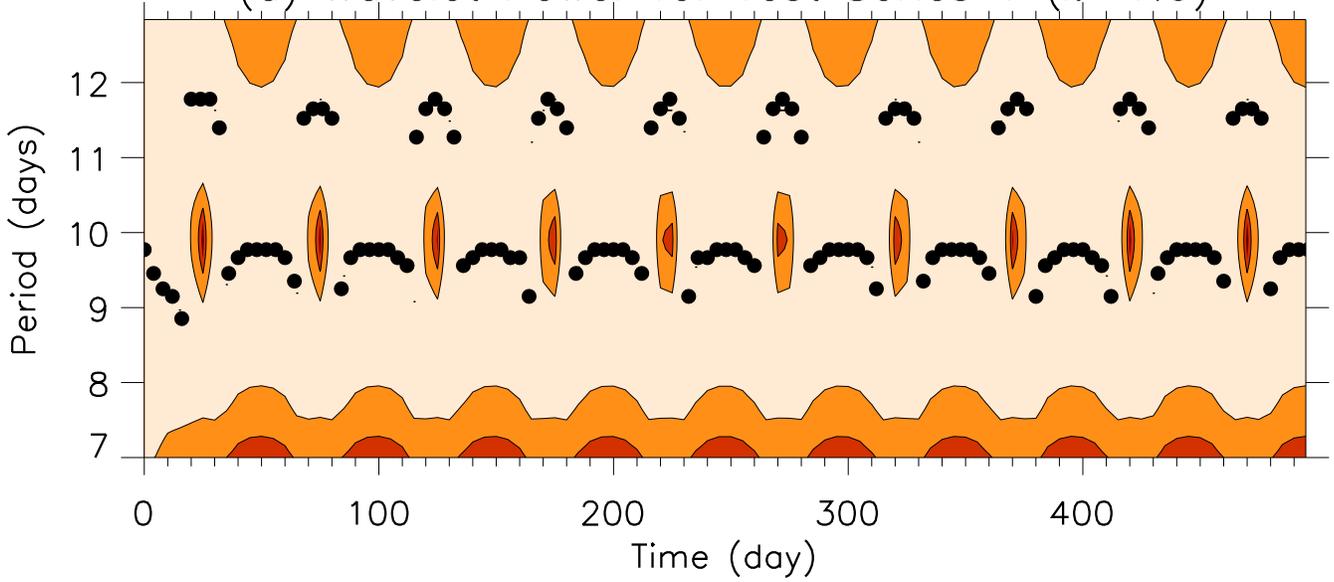
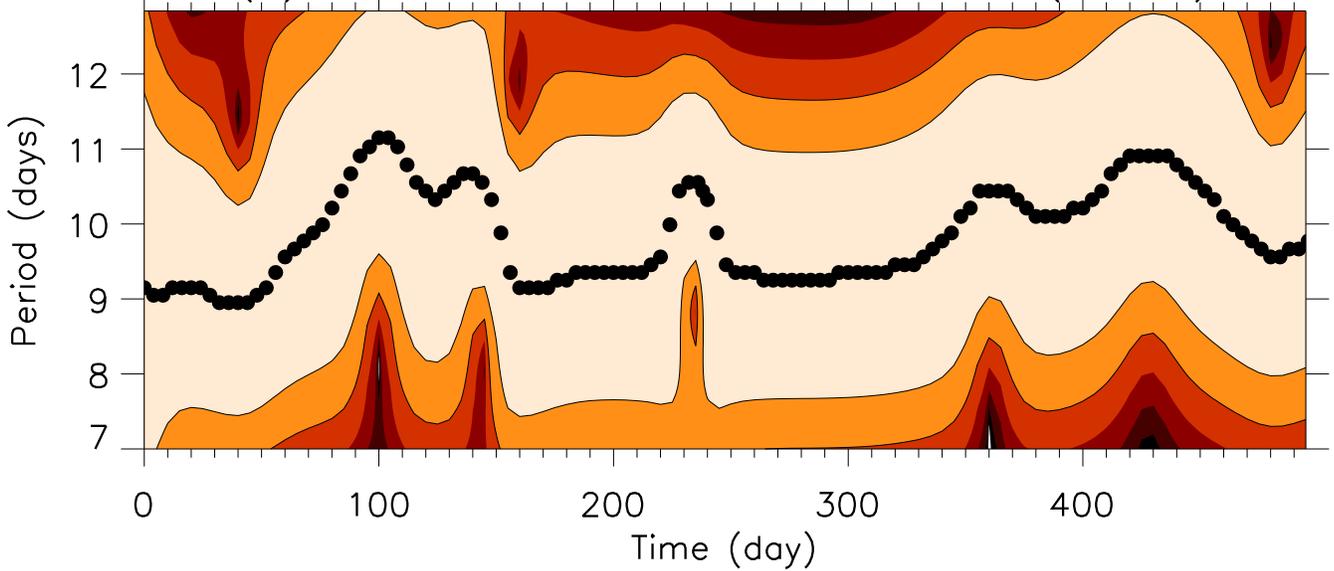
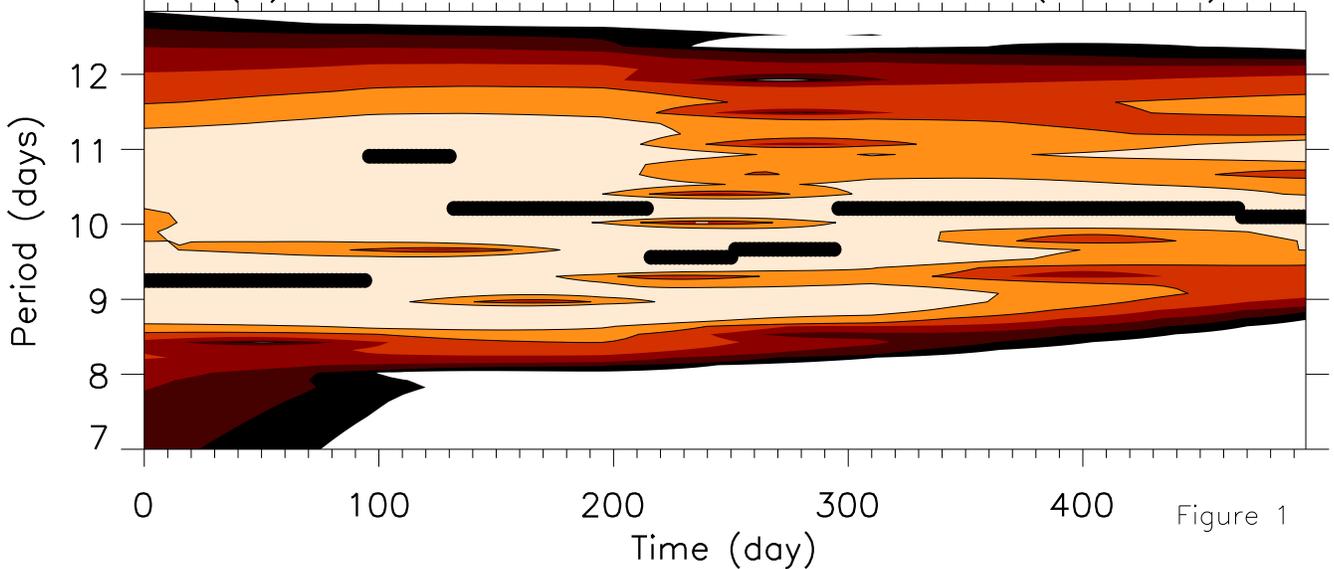
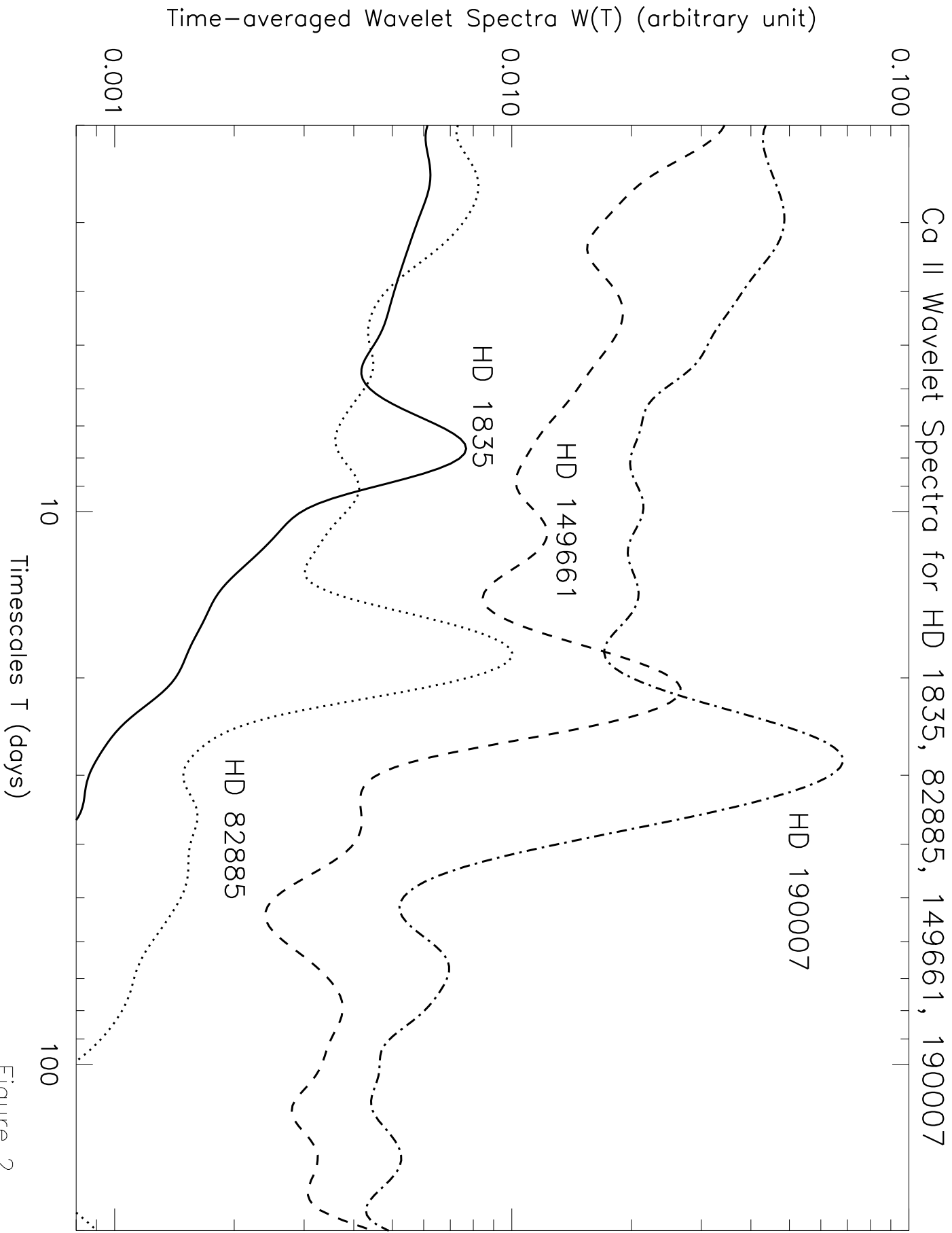
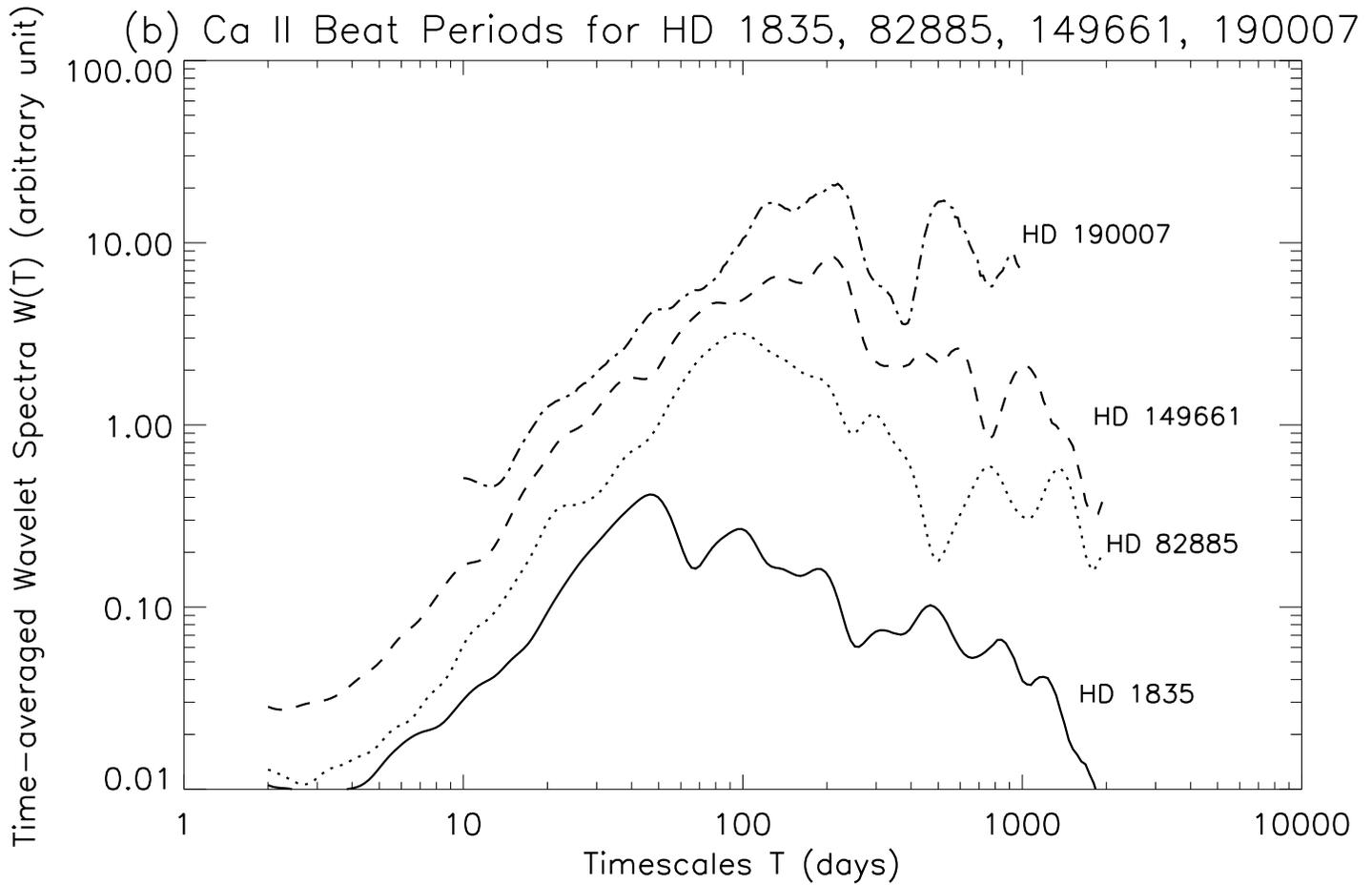
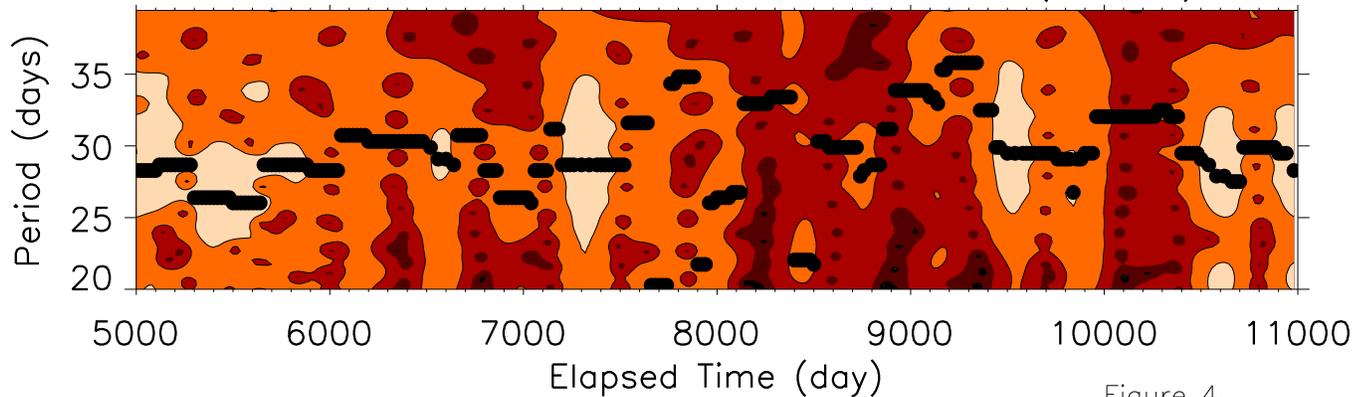
\end{document}